\documentstyle[12pt]{article}
\def\bc{\begin{center}}                \def\ec{\end{center}}
\def\be{\begin{equation}}              \def\ee{\end{equation}}
\def\la{\langle}      \def\ra{\rangle}     \def\l{\left}
\def\r{\right}        \def\dg{\dagger}
\def\alf{\alpha}           \def\lam{\lambda}
\def\Lam{\Lambda}     \def\sig{\sigma}
      \def\pd{\partial}
                     \def\rar{\rightarrow}
\def\hs{\hspace}      \def\vs{\vspace}   \def\pr{\prime}
\def\bt{\begin{tabular}}                  \def\et{\end{tabular}}
\def\sm{\small}    \def\tld{\tilde}     
\def\ld{\ldots}  \def\vphi{\varphi}
\def\matb[#1#2#3#4]{\l(\bt{ll}$#1$&$#2$\\$#3$&$#4$\et\r)}
\def\mata[#1#2]{\l(\bt{l}$#1$\\$#2$\et\r)}

\def\c#1{{\cal #1}}

\begin{document}
\textheight=24cm       \textwidth=16cm   

\begin{flushright}
			Preprint INRNE-TH-95/4 (Aug 1995)\\
                             E-print quant-ph/9801015
\end{flushright}
\medskip

\bc        {\bf CANONICAL EQUIVALENCE OF QUANTUM SYSTEMS,
            MULTIMODE SQUEEZED STATES AND ROBERTSON RELATION}    \ec
\medskip

\centerline{D.A. Trifonov}
\centerline{Institute of nuclear research,}
\centerline{72 Tzarigradsko Chaussee,}
\centerline{1784 Sofia, Bulgaria}
\bigskip \medskip

\begin{abstract}

It is shown that any two Hamiltonians $H(t)$ and $H^\pr(t)$ of $N$
dimensional quantum systems can be related by means of time-dependent
canonical transformations (CT).  The dynamical symmetry group of system
with Hamiltonian $H(t)$ coincides with the invariance group of $H(t)$.
Quadratic Hamiltonians can be diagonalized by means of
linear time-dependent CT.  The diagonalization can be explicitly carried
out in the case of stationary and some nonstationary quadratic $H$.
Linear CT can diagonalize the uncertainty matrix $\sig(\rho)$ for
canonical variables $p_k,\,q_j$ in any state $\rho$, i.e., $\sig(\rho)$ is
symplectically congruent to a diagonal uncertainty matrix.  For multimode
squeezed canonical coherent states (CCS) and squeezed Fock states with
equal photon numbers in each mode $\sig$ is symplectic itself.  It is
proved that the multimode Robertson uncertainty relation is minimized only
in squeezed CCS.

\end{abstract} \vs{1cm}

{\bf 1. Introduction}
\bigskip

The method of  canonical  transformations  (CT)  proved  to  be  a
fruitful approach in treating quantum systems. It is  most  efficient
for  systems  which  are  described  by   Hamiltonians $H$ which are 
quadratic in coordinates  and  momenta,  or  equivalently  in  boson
creation  and annihilation operators (quadratic systems)
[1-6]. The  main advantage of the method of CT consists in reducing the
Hamiltonian $H$  of the treated system $\c S$ to a Hamiltonian $H^\pr $ of
some simple system $\c S^\pr$ with known solutions. 
\newpage \voffset=-2.2cm

Quadratic Hamiltonians $H_{\rm quad}$ model many quantum (and classical)
systems: from free particle and free electromagnetic  field  to  the
waves  in nonlinear  media,  molecular  dynamics  and  gravitational
waveguide [7-10].  A considerable attention to nonstationary quadratic
classical and/or quantum systems is paid in the literature for a long
period  of time (see, for example, references  in  the  review  papers
[9-12]  and recent articles [13-15]. In the last decade they are
intensively  used in quantum optics to describe the nonclassical
properties of light  in squeezed states (SS) [7,16]. In these  (and  many
other)  fields  the nonstationary $H_{\rm quad}(t)$ play important role.

Time dependent CT for quadratic systems $\c S_{\rm quad}$ (classical
and/or quantum)  have been considered in many papers [5,14,17-21]. The
method developed in [5] and intensively used in many papers  (see
references in [10]) consists of construction invariant coordinates  and
momenta $q^0_{j}(t), p^0_{k}(t),\,\, j,k = 1,2,\ld,N,\quad [q^0_{j}(t),
p^0_{k}(t)] = i\hbar
\delta _{kj}$ (or invariant boson operators $A^0_{j}(t), A^{0\dg}_{k}(t)$)
by means  of time  dependent CT.  The generating operator for CT
$p_k,\,q_j \rar p^0_k(t),\,q^0_j(t)$ is  just  the evolution operator of
$\c S_{\rm quad}$. It is worth noting here that the new Hamiltonian $H^\pr$,
when expressed in terms of the invariant coordinates and  momenta
$q^0_{j}, p^0_{k}$ is not diagonal, i.e. such CT are not diagonalizing.

A CT is called diagonalizing if the new Hamiltonian $H^\pr $ in terms
of the coordinates and momenta is diagonal quadratic form with  constant
(and positive) coefficients, i.e. $H^\pr $ is a  Hamiltonian  for  a
system  of uncoupled harmonic oscillators (HO) $H_{\rm ho}$. One has to
distinguish between two different types of diagonalization of $H$:\,\, (a)
when $H^\pr $ is diagonal in terms of  the  new coordinates and momenta
$q_j^\pr , p_k^\pr $ , and \,\,(b) when $H^\pr $ is diagonal  in  terms of
the old variables $q_{j}, p_{k}$. If $H$ is brought to $H^\pr $ by  means
of {\it time-independent} CT, then the spectrums of $H$ and $H^\pr $ are
the same  since $H$ and $H^\pr$  (and the corresponding states) are
unitary equivalent.  If  the relating CT is time-dependent then $H$  and
$H^\pr$  are  no  more  unitary equivalent (but the advantage of the
unitary equivalence  of  the  states remains). So we need a more  general
term of  canonically  equivalent Hamiltonians: $H$ and $H^\pr $ should be
called {\it canonically equivalent} if there is a CT, which transforms $H$
into $H^\pr$.

Not any quadratic boson Hamiltonian can be diagonalized  by  means
of linear time-independent CT [22-24]. Time dependent CT are much more
powerful as we shall see below. Diagonalization of quadratic {\it classical}
Hamiltonian by means of time-dependent CT is studied  by  many  authors
[17-20].  In  [18]  the  general $N$ dimensional classical $H_{\rm
quad}(t)$ has been diagonalized using linear time-dependent  CT.  In
quantum  case   the   diagonalization   problem   (the   second   type
diagonalization) of the one dimensional HO with varying frequency $\omega
(t)$ has been considered recently by Seleznyova [14].

The aim  of  the  present  paper  is  to  consider  the  canonical
equivalence  of $N$ dimensional  quantum  systems,  in  greater  detail
treating the  case  of systems with quadratic Hamiltonians.  As
applications we consider some general  properties  of the dispersion
matrix (called also fluctuation or uncertainty matrix)  and multimode SS.

In  section  II we show   that  any  two $N$ dimensional quantum
Hamiltonians $H(t)$ and $H^\pr$ can be canonically  related  via  time
dependent unitary operator $U(t)$. In the case  of  quadratic
Hamiltonians, $H = H_{\rm quad}(t)$, the operator $U(t)$ is an exponent of a
quadratic form of coordinates $q_{j}$  and momenta $p_{k}$ (that is, a
general element the  methaplectic  group $Mp(N,R)$).  In particular, such
operators can diagonalize any $H_{\rm quad}(t)$.  We perform (in
section  III)  the  diagonalization  of $N$ dimensional $H_{\rm quad}(t)$, 
expressing the parameters of the corresponding linear CT in
terms of solutions of  linear  first  order differential equations. For
$N=1$ these equations are reduced to the equation $\ddot{z} +
\Omega^2(t)z=0$ of classical oscillator with varying frequency. In section
IV the diagonalization of the uncertainty matrix $\sig$ for $p_k,\, q_j$
in any state by means of linear CT is established.  Thus,  $\sig$ is
symplectically congruent to a diagonal dispersion matrix.  For squeezed
canonical coherent states (CCS) [26] and for squeezed Fock states with
equal photon numbers in every mode $\sig$ is symplectic itself.

\bigskip \medskip

\centerline{{\bf 2. Canonical equivalence of quantum systems}}
\medskip

The main aim of the method of CT is  to reduce the Hamiltonian $H$ of the
treated system $\c S$ to a Hamiltonian $H^\pr $  of some simple system $\c
S^\pr$ with known solutions. CT in quantum theory are generated by unitary
operators $U$, which we call the generator of CT.  If CT is {\it
time-independent} then $H$ and $H^\pr $ are unitary equivalent and their
spectrums are the same. However not any pair $H$ and $H^\pr$ can be
related by means of time-independent CT. In particular, not any quadratic
$H$ can be reduced to $H^\pr$ of harmonic oscillator by means of 
time-independent CT [22-24]. The {\it time-dependent} CT are much more
powerful as we shall see below.

Let $|\Psi (t)\ra$ be a solution of the Schr\"odinger equation $[i\hbar
\pd/\pd t - H]|\Psi (t)\ra = 0.$ Then for any unitary operator $U(t)$ the
transformed state $|\Psi ^\pr (t)\ra$,

$$
|\Psi ^\pr (t)\ra = U(t)|\Psi (t)\ra.
\eqno{(1)}$$
 obeys the equation $[i\hbar \pd/\pd t - H^\pr]
|\Psi ^\pr (t)\ra = 0$ with the  new  Hamiltonian $H^\pr $,
$$
H^\pr  = U(t)HU^{\dg}(t) - i\hbar U(t)\pd U^{\dg}(t)/{\pd t}.
 \eqno{(1a)}
$$
 Conversely, if two Hamiltonians $H$ and $H^\pr $ are related
by  means  of  an (unitary) operator $U(t)$  in  accordance  with 
eq. (1a)  then  any solution $|\Psi (t)\ra$ of the system $\c S$ is mapped
into a solution $|\Psi ^\pr (t)\ra$  of the system $\c S^\pr $.
However, not any two given solutions $|\Psi (t)\ra$ and $|\Psi
^\pr (t)\ra$ of the two systems could be mapped into each other by  means
of $U(t)$ since $U(t)$ in general cannot act transitively in the Hilbert
space. A more compact form of relation (1a) is $D^\pr(t) =
U(t)D(t)U^\dg(t)$, where $D(t) = i\hbar\pd/\pd t - H(t)$ ($U(t)$ is
interweaving operator for $D(t)$ and $D^\pr(t)$).
 From the requirement for the mean values of the  "old"  operator
$A$ and the "new" one $A^\pr $,

$$
\la \Psi (t)\l|\matrix{A}\r|\Psi (t)\ra = \la \Psi ^\pr
(t)\l|\matrix{A^\pr }\r|\Psi ^\pr (t)\ra,
\eqno{(2)}$$
 it follows that the operators A and $A^\pr $ are related as

$$
A^\pr  = U(t)AU^{\dg}(t).  \eqno{(2a)}
$$
 Therefore the new canonical operators of the
coordinates  and momenta $q_k^\pr $ and $p_k^\pr , k =1,2,\ld ,N$
are related to the old ones as

$$
q_k^\pr  = U(t)q_{k}U^{\dg}(t), p_k^\pr  = U(t)p_{k}U^{\dg}(t)
.\eqno{(2b)} $$
Two quantum systems should  be  called  {\it canonically  equivalent}  if
there exists unitary operator $U(t)$  which  relates  their  Hamiltonian
operators $H$ and $H^\pr $  in  accordance  with  eq. (1a).  $U(t)$ could
be called $H$-$H^\pr$ {\it canonical equivalence operator} (CEO). Let us
note the main three advantages  of establishing canonical equivalence of
two systems (see also [14], where in fact canonical equivalence of one
dimensional oscillators with constant and time-dependent frequencies was
considered):

(a) If we know solutions for one  of  the  two  canonically  related
systems we can obtain solutions for the other one by means of eq. (1).
For a given system with Hamiltonian $H$ the aim is  to  find  simple
$H^\pr $ with known solutions. As we show below this is always possible:
any $H$ in principle can be brought to as simple $H^\pr $ as that  of
a  system  of stationary oscillators or free particles.

(b) If a time-dependent state $|\Psi ^\pr (t)\ra$ of the  system $\c S^\pr
$  is  an eigenstate  of  an  operator $A^\pr $  then  the
$U(t)$ partner $|\Psi (t)\ra = U^{\dg}(t)|\Psi ^\pr (t)\ra$ in the system
$\c S$ is an eigenstate of the operator $A = U^{\dg}(t)A^\pr U(t)$.

(c) If the operator $A^\pr $ is an integral of motion  for $\c S^\pr $,
that  is $A^\pr$ commutes with the Schr\"odinger operator,

$$
\pd A^\pr /\pd t - (i/\hbar )[A^\pr ,H^\pr ] = 0,\eqno{(3a)}
$$
then the operator
$$
A = U^{\dg}(t)A^\pr U(t)\eqno{(3b)}
$$
 is an integral of motion for the old system $\c S$. This
property is very important  since if we know one solution of a given $\c
S$  we  can  construct  new solutions acting by the invariant 
operators on the known solution [5].

We shall prove now the

{\it Proposition 1}. Any two $N$ dimensional quantum Hamiltonians $H$  and
$H^\pr $ are canonically equivalent. The unitary operator $U(t)$, which
relates $H$ and $H^\pr $ takes the form

$$
U(t) =\hbox{ T\,exp}\l[-\frac{i}{\hbar}\int_{t_0}^{t}
H^\pr(t)dt\r]\,U_{0}\tilde{\rm T}\,\exp\l[{i\over
\hbar }\int_{t_0}^t H(t)dt\r] \equiv
S^\pr(t)U_{0}S^\dg(t),
\eqno{(4)}$$
where $U_{0}$  is  constant  unitary  operator  and $T$  and
$\tilde{T}$  stand  for chronological and  antichronological  product.
The  solution  (4)  is unique for any initial condition $U(0) = U_{0}$.

{\it Proof}. Let us perform two successive time-dependent CT by means of
$U_1=U_{0}S^\dg(t)$ and $U_2=S^\pr(t)$,

$$
S^\dg(t) =\tld{\rm T}\exp\l[{i\over \hbar}\int^{t}H(t)dt\r].\eqno{(4a)}
$$
 Then from eq. (1a) (taking into account $\pd U_1^\dg(t)/\pd t =
(-i/\hbar)HU_1^\dg$) we easily  get $H_{1} = 0$  for  any $U_{0}$.  The
second transformation by means of $U_2=S^\pr(t)$,

$$
S^\pr(t) = {\rm T}\exp\l[-{i\over \hbar}\int^t H^\pr(t)dt\r],
\eqno{(4b)}
$$
 then yields the required result:

$$
H_{2} = UH_{1}U^{\dg} - i\hbar U_{2}\pd U^{\dg}_{2}/\pd t =
-i\hbar S^\pr\pd S^{\pr\,\dg}/\pd t = H^\pr .   \eqno{(4c)}
$$
 Now we see that the direct CT: $H \rar  H^\pr $ is
performed by the operator (4) (note that the product of two unitary
operators  is  also  an  unitary operator).

The definition (1a) of CEO $U(t)$ can be safely multiplied by constant
phase factors. We can prove  that $U(t)$, eq. (4), is uniquely determined
by any initial condition $U(0) = U_{0}$.  Indeed, suppose there is another
unitary operator $\tilde{U}(t)$ which also relates $H$ and $H^\pr $
canonically and $\tilde{U}(0) = U_{0}$.  Now we  note  that  (it is easily
derived from $(1a)$) if $\tilde{U}$ transforms $H$ into $H^\pr $ then
$\tilde{U}^{\dg}$ transforms $H^\pr $ back into $H$ and therefor the
product $V \equiv
\tilde{U}^{\dg}U$ keeps $H$ invariant:
$$
H = VHV^{\dg} + {i\over \hbar}[\pd V/\pd t] V^{\dg}\quad  {\rm
and}\quad V(0) = 1.
$$
The latter can be rewritten  in the form (3a) as $\pd V/\pd t - (i/\hbar
)[V,H] = 0$, which means that $V$ is an integral of motion for the system
$\c S$. Any invariant operator for $H$ has the  form (note that $S(t)$ is
the evolution operator for $\c S$)

$$
V(t) = S(t)V(0)S^\dg(t), \eqno{(5)}
$$
and since $V(0) = 1$ we have $V(t) = S(t)S^\dg(t) =
1.$ In a similar way one can get $U(t)\tilde{U}^{\dg}(t) = 1$ which proves
that the  two  operators $U(t)$  and $\tilde{U}(t)$ coincide. This ends
the proof of proposition 1.

Thus, if initially two vectors $|\Psi ^\pr (0)\ra$ and $|\Psi (0)\ra$
are  related  by means of some $U_{0}$,

$$
|\Psi ^\pr (0)\ra = U_{0}|\Psi (0)\ra.      \eqno{(6)}
$$
 then at later time their evolutions $|\Psi ^\pr (t)\ra$ and
$|\Psi (t)\ra$ governed by $H^\pr $ and $H$ respectively are
canonically related by means of $U(t)$, eq. (4),  with $U(0) = U_{0}$. If
one takes $|\Psi ^\pr (0)\ra = |\Psi (0)\ra$ then $U(0) = 1$  and
one  obtains  unique $U(t)$, eq. (4), which maps isomorphically the time
evolved  state $|\Psi (t)\ra$ of $\c S$ on $|\Psi ^\pr (t)\ra$ of $\c
S^\pr $  for  all  initial $|\Psi (0)\ra$. Note, eq. (6) doesn't imply
that $H(0) = H^\pr (0)$: we  have 

$$ H^\pr (0) = U_{0}HU^{\dg}_{0} -
i\hbar U(0)[\pd U^{\dg}/\pd t]_{|t=0}.$$

Suppose now that $H(t)$ and $H^\pr(t)$ are elements of a Lie algebra $\c
L$.  Then $S \in G \ni S^\pr$, where $G$ is the Lie group generated by $L$
[27]. Thus, the CEO $U(t)\in G$ (for $U_0 = 1$ and for $U_0\in G$ as well)
and one can use the known properties of $G$ [27] to represent $U(t)$ in
other factorized forms.

The operator (4) converts canonically any $N$ dimensional $H$ into  any
desired $N$ dimensional $H^\pr $. In particular $H$ can be converted
into $H^\pr$ for a system of $N$ free particles or for a  system  of
uncoupled  harmonic oscillators ($N$ mode free boson field). In the latter
case  if $H$  is  a quadratic form in terms  of $N$  canonical  operators
$q_{k}$  and $p_{j}$  the operator  (4)  solves  the  {\it diagonalization
problem} for quadratic Hamiltonians.

We stress that one has  to distinguish between two type of diagonalization:
\, (a) $H^\pr$ is diagonal in terms  of  the  new variables $q_k^\pr\,
p_k^\pr $, and \, (b) $H^\pr$ is diagonal in terms of the old
variables $q_{k}\, p_{k}$.  In the  first  case  the two  systems
$\c S$  and $\c S^\pr $ respectively are treated in two
different ($q$-  and $q^\pr$- ) coordinate representations (wave
functions $\Psi (q,t) = \la q|\Psi(t) \ra$ and $\Psi^\pr(q^\pr,t) =
\la q^\pr |\Psi^\pr(t)\ra)$,  while  in  the second case we can work in
the same $q$-representation  (wave  functions $\Psi (q,t) = \la q|\Psi
(t)\ra$ and $\Psi ^\pr (q,t) = \la q|\Psi ^\pr (t)\ra$). The second
kind diagonalization is achieved by means of operator $U(t)$, eq. (4), with
$H^\pr$ of the form of Hamiltonian $H_{\rm ho}$ of $N$ attractive
oscillators,

$$ H^\pr  = {1\over 2}\sum_{k}^{N}\l[{1\over
m_{k}}p_k^{2}+ m_{k}\omega^{2}_{k}q_k^{2}\r] \equiv
H_{\rm ho}, \eqno{(7)} $$

Let us briefly discuss the two CT, generated by  $U_1= U_{0}S^\dg(t)$ and
$U_{2}(t)=S^\pr(t)$ (see proposition 1). The first one brings $H$ to zero,
therefore the  new  states $|\Psi \ra_{1}$  are  time-independent.
This is because $S^\dg(t)$ is an evolution  operator for the
$\c S$ backward in time. After this CT the new canonical variables

$$q^{(1)}_{k}= U_{1}(t)q_{k}U^{\dg}_{1}(t)\quad {\rm and}\quad
p^{(1)}_{j}= U_{1}(t)p_{j}U^{\dg}_{1}(t)$$
obey  the equations ($\pd U_1/\pd t
= iU_{1}H$, $\pd U_1^\dg/\pd t = -iHU_1^{\dg}$)

$$
\frac{\pd q^{(1)}_{k}}{\pd t} = {i\over \hbar}[U_1HU^\dg_1,q^{(1)}_{k}],
\quad
\frac{\pd p^{(1)}_{k}}{\pd t} = {i\over \hbar} [U_1HU^\dg_1,p^{(1)}_{k}],
\eqno{(8b)}
$$
i.e., $q^{(1)}_k,\,p^{(1)}_k$ are Heisenberg operators for the system 
$\c S $. The operator $S^\pr(t)$  is  the evolution operator
forward in time for the target system $\c S^\pr$.

It is worth noting here the CT generated by the evolution operator $S(t)$
of $\c S$.  This CT converts $H(t)$ into $H^{\pr\pr}(t) =
S(t)H(t)S^{\dg}(t) + H(t)$.  If $H$ is time-independent then $H^{\pr\pr}$
= 2H. From (2) we derive that the new canonical variables after CT (9a),

$$
q^{\pr\pr}_k = S(t)q_{k}S^{\dg}(t)\equiv q^0_k,\quad
p^{\pr\pr}_k = S(t)p_{k}S^{\dg}(t) \equiv p^0_k,
\eqno{(9a)} $$
when expressed in terms of the old ones $q_{k},\,p_{j}$, are {\it
integrals of motion} of $\c S$, satisfying the  eqs.  (6a,b).  Then  the
operators

$$
q^0{}^\pr_{k} = U(t)q^0_kU^{\dg}(t), \quad p^0{}^\pr_k =
U(t)p^0_kU^{\dg}(t)        \eqno{(9b)}
$$
 where $U(t)$ is given by eq. (4), are integrals of motion for
the system $H^\pr$  (recall the above property (c)  of  the canonical
equivalence).  Such  integrals  of  motions  for  quadratic systems $H(t)$
have been constructed in [5] and intensively  used  later [10].

If the target Hamiltonian $H^\pr $ is a sum of  stationary
attractive harmonic oscillators $H_{\rm ho}$, eq. (7), in terms (for
example) of the intermediate variables $q^{(1)}_{k},\,p^{(1)}_{k}$. Then 
the second CT (generated by $U_{2}(t) = \exp(-(i/\hbar)H_{\rm ho}t)=
S_{\rm ho}(t)$):
$(q^{(1)}_{k},\, p^{(1)}_{k}) \rar  (q_k^\pr, p_k^\pr)$ yields the
explicit relations  between  $q^{(1)}_{k},\,p^{(1)}_{k}$ and the final
canonical variables $q^\pr, p^\pr$,

$$
q_k^\pr  = q^{(1)}_{k}\cos (\omega _{k}t) + \frac{1}{m_{k}\omega
_{k}}p^{(1)}_{k}\sin (\omega _{k}t),    \eqno{(10a)}
$$
$$
p_k^\pr  = -m_{k}\omega _{k}q^{(1)}_{k}\sin(\omega_{k}t) +
p^{(1)}_{k}\cos(\omega_{k}t).             \eqno{(10b)}
$$
 This  is  an  orthogonal  symplectic transformation  which  for
every $k$  is  a  rotation  in  the   plane $q^{(1)}_{k},(m_{k}\omega
_{k})^{-1}p^{(1)}_{k})$.  It is evident that the CT (9a)
 {\it is not}  a diagonalizing CT, unless $H$ itself is of the form
$H_{\rm ho}$.

It is of practical interest for a given Hamiltonian $H$ to specify the
class of CT (possibly time-dependent) which  keep  the $H$ invariant (and
thus they keep  the  Schr\"odinger  equation invariant), i.e.,

$$
H^\pr \equiv U(t)HU^\dg(t) -i\hbar U(t){\pd U^\dg(t) \over\pd t}  = H.
\eqno{(11)}
$$
{} For time-independent $U$ eq. (11) reduces to $HU = UH$. From (4) and
(11) we have  $U(t) = S(t)U_0S^\dg(t)$, where $U_0$ is arbitrary unitary
operator. Thus, $H$-$H$ CEO are unitary invariant operators.
In the first  paper  of  refs. [5] the dynamical symmetry group of a
system $\c S$ has been defined as a group of unitary operators which commute
with $i\hbar\pd/\pd t - H$ and act irreducibly in the Hilbert space. Now
we see that this symmetry group is highly nonunique since $U_0$ is
arbitrary and one can take it from irreducible representations of many
groups. Then the set of invariants $S(t)U_0S^\dg(t)$ realize an equivalent
representation.  For example, by means of the invariants $q^0_{k}$  and
$p^0_{k}$ one can construct  an irreducible representation of (the Lie
algebra of)  the Heisenberg-Weyl group $H_W(N)$ and the quasi unitary
group $SU(N,1)$ as well [5,27].  This  means  that the groups $H_W(N)$
and $SU(N,1)$ can  be considered on equal as dynamical symmetry groups of
any $N$ dimensional system.

In the next section we consider the above  described  canonical
equivalence approach  in greater detail for quadratic quantum systems, for
which some  explicit solutions can be obtained.

\medskip\bigskip

\centerline{{\bf 3. Canonical transformations of quadratic systems}}
\centerline{{\bf and diagonalization.}}
\medskip

We consider the general $N$ dimensional  nonstationary  quantum system
with Hmiltonian $H(t)$ which is a homogeneous quadratic form of
coordinates and momenta (in this section $H_{\rm quad} \equiv H$), 

$$
H(t) = \c A_{jk}(t)p_{j}p_{k}+ \c B_{jk}(t)p_{j}q_{k}+ \tld\c B
_{jk}(t)q_{j}p_{k} + \c C_{jk}(t)q_{j}q_{k},   \eqno{(12a)}
$$
 where the coefficients $\c A_{jk}(t) = \c A_{kj}(t)$,
$\c B_{jk}(t),\,\tld\c B_{jk}(t)$   and $\c C_{jk}(t) =  \c
C_{kj}(t)$ are arbitrary real functions of time. By no lack of generality
one can put $\c B_{jk}(t) = \tld\c B_{kj}(t)$. In (12a) the
summation over the repeated  indices is adopted. We can introduce
$N$ component vectors

$$
\vec{q} = (q_{1},q_{2},\ld
,q_{N}),\quad \vec{p}= (p_{1},p_{2},\ld ,p_{N})
$$
 and $N\times N$ real matrices $\c A(t)$, $\c B(t)$,
$\c C(t)$ ($\c A(t)$ and  $\c C(t)$  are  symmetric)  and
rewrite the Hamiltonian (12a) in a more compact form

$$H(t) = \vec{p}\c A(t)\vec{p} + \vec{p}\c B(t)\vec{q} +
\vec{q}\c B(t)^T\vec{p} + \vec{q}\c C(t)\vec{q}, $$

 where $\c B^{T}$ is the transposed of $\c B$. To shorthand the notations
it is convenient to introduce the $2N$-vector $\vec{Q} =
(\vec{p},\vec{q})$ and $2N\times 2N$ matrix $\c H$ (the grand matrix) and
rewrite the Hamiltonian (12a) as ($\mu ,\nu  = 1,2,
\ld, 2N$)
$$
H(t) = \c H_{\mu \nu }Q_{\mu }Q_{\nu } \equiv \vec{Q}\c
H(t)\vec{Q},     \eqno{(12b)}
$$
 where $\c H =
\matb[{\c A} {\c B} {\c B^{T}} {\c C}]$.

We note that nonhomogeneous quadratic Hamiltonians  (i.e., Hamiltonians
(12a,b) with linear terms added)  can be easily  reduced  to  the  form
(12a,b)   by   means   of   simple   time    dependent    displacement
transformations.

Let $H^\pr $ be an other quadratic Hamiltonian
$$
H^\pr (t) =\vec{Q}\c H^\pr (t)\vec{Q}, \c H^\pr  =
\matb[{\c A^\pr} {\c B^\pr} {\c B^\pr{}^{T}} {\c C^\pr}].
\eqno{(13)}$$
 Then the unitary operator $U(t)$, eq. (4), which relates
canonically  Hamiltonians (12) and (13), is an exponent of a quadratic in
$\vec{q}$ and $\vec{p}$ form (we take $U_0\in Mp(N,R)$),
$$
U(t) = S^\pr(t)U_0S^\dg(t) = \exp\l[{i\over \hbar }\vec{Q}\tld{\c H}(t)
\vec{Q}\r],     \eqno{(14)}
$$
 where $\tilde{\c H}(t)$ is a new grand matrix of the form
(12b) and (13). $\tilde{\c H}(t)$  can be expressed in terms of the
Hamiltonian matrices $\c H(t)$ and $\c H^\pr (t)$ using the
Baker-Campbell-Hausdorff formula. In this case the  operator  (14)
generates linear transformation of coordinates and momenta  (we  write it
in $N\times N$ and $2N\times 2N$ matrix forms),

$$\vec{Q}^\pr = \Lam(t)\vec{Q}\quad {\rm or}\quad \mata[{\vec{p}^\pr}
{\vec{q}^\pr}] =
\matb[{\lam_{pp}} {\lam_{pq}} {\lam_{qp}} {\lam_{qq}}]\mata[{\vec{p}}
{\vec{q}}], \eqno{(15)}
$$
where $\lam_{pp}, \lam_{pq}, \lam_{qp} \hbox{ and } \lam_{qq}$ are
$N\times N$ matrices,
$$
\Lam  = \matb[{\lam_{pp}} {\lam_{pq}} {\lam_{qp}} {\lam_{qq}}].
\eqno{(15a)} $$

 {}From eqs. (1a), (12b), (13) and (15) we obtain the following
relation between the symmetric matrices $\c H$, $\c H^\pr$ and
$\tilde{\c H}$ (14) and the symplectic matrix $\Lam $,

$$
{d\over dt}\tilde{\c H}(t) = -\c H^\pr (t) + \Lam ^{T}\c H(t)\Lam .
\eqno{(16)}$$
 We see that for a given $\tilde{\c H}(t)$ and $\c H(t)$
this is a simple linear equation for $\c H^\pr (t)$. However for a
given Hamiltonian matrices $\c H$ and $\c H^\pr $  this  is
highly nonlinear equation for $\tilde{\c H}(t)$ since the  matrix
$\Lam (t)$  is  to  be expressed in terms of $\tilde{\c H}(t)$
again: $\Lam \vec{Q}= U(t)\vec{Q}U^{\dg}(t)$.  Nevertheless  for any given
(differentiable with respect to $t$) matrices $\c H(t)$  and $\c
H^\pr (t))$ and for a given  initial  condition $\tilde{\c H}_{0}$
the  above  equation  has  unique solution for $\tilde{\c H}(t)$,
since the expression of $\Lam $ in terms  of $\tilde{\c H}$  is
also differentiable and Peano theorem could be applied [37].

In this scheme $\Lam (t)$ is naturally represented as a  product  of
two other $2N\times 2N$ matrices $\Lam ^{(1)}$ and $\Lam ^{(2)}$ of
the form (15)  corresponding  to the two successive CT generated by
$U_1(t)$ and $U_2(t)$:
$$
\Lam  = \Lam ^{(2)}\Lam ^{(1)};\quad \vec{Q}^{(1)} =
\Lam^{(1)}\vec{Q},\quad \vec{Q}^\pr  = \Lam ^{(2)}\vec{Q}^{(1)}.
\eqno{(17)}
$$
 The matrices $\Lam ^{(1)}$ and $\Lam ^{(2)}$ are seen to
be solutions of  the  first  order  linear equations,
$$
{d\over dt}\Lam ^{(1)}= \Lam^{(1)}F^{(1)}(t),\quad {d\over
dt}\Lam ^{(2)}= F^{(2)}(t)\Lam^{(2)},
\eqno{(18)}$$
 where
$$
F^{(1)}(t) = -2J\c H(t), \quad F^{(2)}(t) = 2J\c H^\pr (t),\,\, J =
\matb[0 1 {-1} 0],        \eqno{(18a)}
$$
 or in terms of the $N\times N$ matrices $\c A$, $\c B$,
$\c C$  and $\c A^\pr$, $\c B^\pr$, $\c C^\pr $,
$$
F^{(1)}(t) = 2 \matb[{-\c B^{T}} {-\c C} {\c A} {\c B}],
\qquad F^{(2)}(t) = -2 \matb[{-\c B^\pr{}^{T}} {-\c C^\pr} {\c
A^\pr} {\c B^\pr}] .      \eqno{(18b)}
$$
 If $H^\pr$ is diagonal as for the oscillator system (7)  then
the  second  eq. (18)  is  easily
solved: $\Lam^{(2}(t) = \exp(2J\c H_{\rm ho} t)\Lam^{(2)}_0$.
To perform the diagonalization of a quadratic $H$ one has also to solve
the first matrix equation in (18) and obtain $\Lam ^{(1)}(t)$, which in
principle is always possible.  In  the case  of stationary initial $H$ the
$\tilde{T}$ exponent becomes ordinary  one,  so the  explicit solution  is
given by the matrix exponent $\Lam^{(1)}_0\exp(-2J\c H t)$.  The new
variables are
stationary $H$ the diagonalizing CT is (highly nonunique)
$$\vec{Q}^\pr = \exp(2J\c H_{\rm ho} t)\Lam^{(2)}_0\Lam^{(1)}_0\exp(-2J\c
H)\vec{Q} .
\eqno{(18c)}
$$
One can put $\Lam^{(1,2)}_0=1$, which corresponds to $U_0$ in eq. (4).
Having  obtained explicitly $\Lam (t) = \Lam^{(2)}\Lam ^{(1)}$ one
can next try to solve eq. (16) and obtain the generating operator $U(t)$
in the form of the quadratic exponent (14).

Note, the resulting $H^\pr $ is diagonal in the variables, which we choose
for $H_{\rm ho}$. Let those variables be $p^{(1)}_k,\,q^{(1)}_k$. Then the
final variables $p^\pr_k,\, q^\pr_k$ obey eqs. (10a,b). Inverting the
transformations (10a,b) we obtain $H^\pr$ diagonal in terms of the final
variables as well: $H^\pr = H_{\rm ho}(\vec{p}^\pr,\vec{q}^\pr)$. In this
way we perform explicitly the first kind diagonalization. If $H^\pr =
H_{\rm ho}$ in terms of old variables $p_k,\, q_k$ (second kind
diagonalization), then  $H^\pr$ is evidently not diagonal in terms of
$p^\pr_k,\, q^\pr_k$.

{}For some time-dependent $H(t)$ explicit solutions of eqs. (18) can also
be found.  Thus, in the case of $N=1$ following the scheme of refs. [4,5]
one can express matrix elements of $\Lam^{(1)}(t)$ (and therefore the CEO
$U(t)$ and all quantum mechanical solutions of the system $\c S$) in terms
of a complex function $z(t)$ which obeys the equation of classical
oscillator 

$$\ddot{z} + \Omega^2(t)z =0, $$ 
where $\Omega^2(t)$ is simply determined by the parameters of the
Hamiltonian (12),

$$\Omega^2(t) = 4\c
A\c C + 2\c B\dot{\c A}/\c A + \ddot{\c A}/(2\c A) - 3\dot{\c A}^2/(4\c
A^2) - 4\c B^2 - 2\dot{\c B}.$$ 
For HO with varying frequency $\omega(t)$ we have $\Omega^2(t) =
\omega^2(t)$. It is seen that an $\Omega(t)$ corresponds to a class of
quadratic $H(t)$. For example constant $\Omega$ corresponds to the
stationary oscillator and to the oscillators with varying mass (damped
oscillators) $m(t) = m_0\exp(-2bt)$ and $m(t)=m_0\cos^2bt$ considered in
[10,19]. Analytical solutions to the equation of $z(t)$ are known for a
variety of "frequencies" $\Omega(t)$ [37]. In the case of an oscillator
with varying frequency the diagonalizing CT generator $U(t)$ has been
expressed in terms of $z(t)$ in [14].

Let us briefly  discuss  the  algebraic  aspects  of  the  above
results. From the canonical commutation relations it follows that  all
$2N\times 2N$ matrices $\Lam $ obey the relation (the symplectic
conditions)
$$
\Lam J \Lam ^{T}= J,         \eqno{(19a)}
$$
 which for the $N\times N$ matrices $\lambda _{qq}, \lambda
_{pp}, \lambda _{qp}$ and $\lambda _{pq}$, defined in eq. (15)),  read
$$
\lam_{qq}\lam^T_{pp} - \lam_{qp}\lam^{T}_{pq} = 1, \quad
\lam_{qq}\lam^T_{qp} = \lam_{qp}\lam^{T}_{qq},\quad \lam_{pq} \lam^T_{pp}
= \lam_{pp}\lam^T_{pq} .         \eqno{(19b)}
$$
 The set of matrices which obey the relation (19a) is  defined as the
symplectic matrix group $Sp(N,R)$. It has $N(2N+1)$ real  parameters.  The
rank of its Lie algebra is $N$.\, It is known that in classical mechanics the
set of linear homogeneous CT forms  a symplectic group $Sp(N,R)$.  In  the
quantum  case  the  set  of  matrices $\Lam $,  which  realize homogeneous
linear transformations of the operators of coordinates and momenta close
the same group.  However  the  set  of unitary operators $U$ for which
$U\vec{q}U^{\dg}$ and $U\vec{p}U^{\dg}$ are linear combinations of
$\vec{p}$ and $\vec{q}$ contains one extra parameter, namely the phase
factor.  If  one considers CT in  greater  detail  as  transformations of
coordinates, momenta and {\it vectors} in Hilbert  space  one  has  to
count  the  phase factors as well and then we  get  the  larger  group
$Sp(N,R)\times U(1) \equiv \tilde{Mp}(N,R)$. If we consider
transformations of coordinates,  momenta and {\it states} we have to
factorize over $U(1)$: $\tilde{Mp}(N,R)/U(1) = Mp(N,R)$.  The resulting
group $Mp(N,R)$ is called {\it methaplectic} group.  It  is double covering
of $Sp(N,R)$.  The  Lie algebras of $Mp(N,R)$ and $Sp(N,R)$ are isomorphic
[25].  They  are  of dimensions $N(2N+1)$ and  this  is the number  of
independent  matrix elements of matrix $\tilde{\c H}$ in (14). The
generators $U(t)$ of linear CT (15) can be considered as operators of  the
unitary  (but  not  faithful) representation $U(\Lam )$ of the symplectic
group $Sp(N,R)$. One can use the group representation technique [27] to
represent $U(t)\in Sp(N,R)$ in several factorized forms.  In the case of
one dimensional nonstationary HO the CEO $U(t)\in SU(1,1)$ and its 
factorized forms have been considered in [14].

If one considers Hamiltonians  (12)  with  linear  terms
$\vec{d}(t)\vec{p} +  \vec{e}(t)\vec{q}$  added, then  in  the  same  way
one  would  get  that  such inhomogeneous quadratic Hamiltonians can be
diagonalized to  the  form (7) by means of the same $U(t)$, eq. (4), this
time $U(t)$ being an element of the semidirect product group
$Mp(N,R)\times\!\!\!\!\!\supset H_{w}(N)$, where $H_{w}(N)$  is  the
$N$ dimensional   Heisenberg-Weyl   group.

\bigskip\medskip

\centerline{{\bf 4. Diagonalization of the dispersion matrix and}}
\centerline{{\bf multimode squeezed states}}
\medskip

The established possibility  of  converting (by means of time-dependent
CT) any $N$ dimensional Hamiltonian $H$  to $H_{\rm ho}$ of harmonic
oscillators suggests to expect  that the dispersion matrix
$\sig(\vec{Q},\rho )$ of canonical observables $Q_\nu$, $\nu =1,2,\ld 2N$
in any (generally mixed) quantum state $\rho $ could be diagonalized by
means of some  state dependent  CT.  It turns out that this really holds.

Let  us recall the notion of dispersion matrix $\sig(\vec{Q},\rho)$
(called also fluctuation matrix, or uncertainty matrix). This is an
$2N\times 2N$ matrix constructed by means  of  the  second  momenta (the
dispersions or the variances and covariances) of coordinates $q_{k}$ and
momenta $p_{l}$ ($k,\,l  = 1,2, \ld , N; \quad Q_{k} = p_{k}, Q_{N+k} =
q_{k}$)

$$
\sig_{\mu \nu } \equiv  \sig_{Q_{\mu}Q_{\nu}} = {1\over 2}\la Q_{\mu}
Q_{\nu}+ Q_{\nu}Q_{\mu}\ra - \la Q_{\mu}\ra \la Q_{\nu}\ra,
\eqno{(20)}
$$
 where $\la Q\ra$ is the mean value of $Q$: for pure states $|
\Psi \ra$ we write $\la Q\ra =
\la\Psi \l|\matrix{Q}\r|\Psi \ra$ and for mixed state $\rho $
 the mean values are $\la Q\ra = {\rm Tr}\,(Q\rho )$. The  matrix $\sig$  is
symmetric by construction. If we do not need to specify pure or  mixed is
the state (or the concrete choice of variable $Q$)  we will simply  write
$\sig(Q)$ (or $\sig(\rho)$).

{} For pure states $|\Psi \ra$, which  are  unitary  equivalent  either to
multimode CCS $|\vec{\alf}\ra$ ($\vec{\alf} =
(\alf_{1},\alf_{2}, \ld , \alf_{N})$, $N = 1,2,\ld$), or to Fock states
$|\vec{n}\ra$ ($\vec{n}= (n_1,n_2,\ld,n_N)$)
$$
|\Psi \ra = U_{0}|\vec{\alf}\ra,\quad{\rm or}\quad |\Psi\ra=U_0|\vec{n}\ra
\eqno{(21)}$$
 the dispersion matrix $\sig(\vec{Q},\Psi )$ can always be
diagonalized by some CT. To show this let us note that for any CT,
generated by some unitary $U$,

$$\vec{Q} \rar \vec{Q}^\pr  = U\vec{Q}U^{\dg}\qquad (\vec{Q} =
(\vec{p},\vec{q})), \eqno{(22)}$$
 we have (using eqs. (20), (2) and (2a)) the following relation
for the dispersion matrices,
$$
\sig(Q^\pr ,\rho ) = \sig(\vec{Q},\rho ^\pr ),\qquad
\sig(Q^\pr ,\Psi ) = \sig(\vec{Q},\Psi ^\pr ),\eqno{(23)}
$$
where the new state is $\rho ^\pr  = U\rho U^{\dg}$ and the new pure
state is $|\Psi ^\pr \ra = U|\Psi \ra$. Taking $U = U^{\dg}_{0}$ and
$|\vec{\alf}\ra = |\Psi ^\pr \ra$ (or $|\vec{n}\ra = |\Psi ^\pr
\ra$)  in eq. (23b) we obtain 

$$
\sig(Q^\pr,\Psi ) = \sig(\vec{Q},\vec{\alf})\quad{\rm or}\quad
\sig(Q^\pr,\Psi ) = \sig(\vec{Q},\vec{n}).  \eqno{(23b)}
$$
 The dispersion matrices $\sig(\vec{Q},\vec{\alf})$ and $\sig(\vec{Q},
\vec{n})$ are diagonal. Therefore,  $\sig(Q^\pr,\Psi )$ is also diagonal.
This ends the proof that for any state $|\Psi \ra$ which obey eq. (21) 
the dispersion  matrix $\sig(U\vec{Q}U^\dg ,\Psi )$ is diagonal.  Note  that
$\sig(\vec{Q},\vec{\alf})$ has the special diagonal form:
$$
\sig_{11}(\vec{Q},\vec{\alf}) = \sig_{22}(\vec{Q},\vec{\alf}) = \ld
 = \sig_{NN}(\vec{Q},\vec{\alf}),  \eqno{(24a)}
$$
$$
\sig_{N+1,N+1}(\vec{Q},\vec{\alf}) = \sig_{N+2,N+2}(\vec{Q},\vec{\alf}) = \ld
 = \sig_{2N,2N}(\vec{Q},\vec{\alf}),  \eqno{(24b)}
$$
$$
\det\sig(\vec{Q},\vec{alf}) =
\prod_k^N\sig_{kk}(\vec{Q},\vec{\alf})\cdot\sig_{N+k,N+k}(\vec{Q},
\vec{\alf}) = ({\hbar^2 \over 4})^{N}.   \eqno{(24c)}
$$
If $|\Psi (t)\ra$ is a time evolved state for a system $\c S$ and
$|\Psi(0)\ra$ satisfies the first equality in (21),  then  the time
dependent CT (2) generated by $U(t)$, eq. (4),  with $H^\pr  = H_{\rm
ho}(\vec{p},\vec{q})$, and $U_{0}$ taken  as in  eq.  (21),  will
preserve in time  the diagonal  form  (24)   of $\sig(Q^\pr (t),\Psi (t))$
since the evolution of CCS $|\vec{\alf}\ra$ governed  by $H_{\rm ho}$ is
stable (up to a phase factor): $|\vec{\alf}\ra \rar |\vec{\alf(t)}\ra$.

All group-related CS $|\Psi (g)\ra$ [26,27]  with  initial (reference)
vector $|\Psi _{0}\ra = |\alf\ra$ or $|\Psi _{0}\ra = |\vec{n}\ra$  (and
any group $G\ni g$) by  construction are  unitary equivalent  to
$|\vec{\alf}\ra$ (or to $|\vec{n}\ra$) , $|\Psi(g)\ra = U_0(g)\Psi_0\ra$,
and thus their  dispersion matrix  obeys eq. (21), i.e.
$\sig(\vec{Q},\Psi (g))$  is  diagonalized  to  the  form  (24) by  means
of the (nonlinear) CT  $\vec{Q}\rar \vec{Q}^\pr  = U_{0}(g)\vec{Q}
U^{\dg}_{0}(g)$.

From this construction it is however not clear  whether  one  can
diagonalize $\sig(\vec{Q},\Psi )$ for any state $|\Psi \ra$  since  in
general $|\Psi \ra$  is  not unitary equivalent neither to CCS
$|\vec{\alf}\ra$ nor to Fock state $|\vec{n}\ra$.  Surprisingly,  it
turned  out  that $\sig(\vec{Q},\rho )$ for arbitrary state $\rho $
{\it can always be diagonalized by means of linear CT}. The crucial point
is the  positive definiteness  of  the dispersion matrix which is
established by  the  following  proposition (some known properties  of
the dispersion  matrix  are  reviewed in, for example, ref. [10]).

{\it Proposition 2}. The dispersion matrix $\sig(\vec{Q},\rho )$ is
positive  definite for any (pure or mixed) state $\rho$.

{\it Proof}. Consider the mean values of the positive definite
$\vec{z}$ family of operators $F^{\dg}(\vec{z})F(\vec{z})$,
$$
F(\vec{z}) = \vec{z}\Delta \vec{Q}\equiv \sum_\nu^{2N}z_{\nu}
\Delta Q_{\nu},\quad  \Delta Q_{\nu} = Q_{\nu}- \la Q_{\nu}\ra,
\eqno{(25b)}
$$
 $z_{\nu}$ being arbitrary complex  numbers  ($\vec{z}$  is  an
$2N$ component  complex vector). In any state $\rho $ one has
$$
Tr[\rho F^{\dg}(\vec{z})F(\vec{z})] \equiv  \la
F^{\dg}(\vec{z})F(\vec{z})\ra \ge  0 .
\eqno{(26)}
$$
 The mean value $\la F^{\dg}(\vec{z})F(\vec{z})\ra$ is expressed
in  terms  of  the  dispersion matrix (20) as
$$
\la F^{\dg}(\vec{z})F(\vec{z})\ra = \vec{z}^{*}[\sig(\vec{Q},\rho ) - {i\over
2}\hbar J]\vec{z}\equiv \vec{ \vec{z}}^{*}\Phi \vec{\vec{z} }\ge  0,
\eqno{(26a)}
$$
 where $J$ is given by eq. (18a) and $\vec{z}^{*}J\vec{z} =
\sum_{\mu,\nu}z^{*}_{\mu} J_{\mu\nu}z_{\nu }$. The
$2N\times 2N$ matrix $\Phi $ is nonnegative definite in virture of eq.
(26).  It is known [10] that the dispersion matrix for canonical
observables $\sig(\vec{Q},\rho)$ is also nonnegative  definite  and obeys
the Robertson inequality [28]

$$
\det \sig(\vec{Q},\rho) \geq  ({\hbar^{2}\over 4})^{N}.
\eqno{(27)}
$$
 Now we shall prove that $\sig(\vec{Q},\rho )$ is {\it positive
definite}.  In  this purpose we put $\vec{z} = \vec{x} + i\vec{y}$,
$\vec{x}$ and $\vec{y}$ being arbitrary real vectors and write
$$
0 \le \vec{z}^{*}[\sig(\vec{Q},\rho) - {i\over 2}\hbar J]\vec{z} =
\vec{x}\sig(\vec{Q},\rho)\vec{x} + \vec{y}\sig(\vec{Q},\rho) \vec{y} -
\hbar \vec{x}J\vec{y}.       \eqno{(28a)}
$$
 If we put $\vec{y} = 0$ in (28a) keeping $\vec{x}$ arbitrary we
would have $\vec{x}\sig(\vec{Q},\rho )\vec{x} \ge
0$, i.e. the dispersion matrix is nonnegative definite.  From  [29]  we
know that if a matrix $\sig$ is nonnegative defined and its determinant  is
strictly positive, then $\sig$ is positive definite.  Thus,  the  dispersion
matrix is positive definite, $\sig(\vec{Q},\rho )\vec{x} > 0$. End of
the proof.

Several interesting properties of  the  dispersion  matrix
$\sig(\vec{Q},\rho )$ follow from its positive definiteness. The first to
note is related to its diagonalization. A well established fact is that
any symmetric and positive definite $2N\times 2N$ matrix $\sig$  can  be
diagonalized  by  means  of symmetric transformation with a symplectic
matrix $\Lam$  [22-24]: $\Lam \sig\Lam ^{T} =
{\rm diag}\{d_{1},d_{2},\ld ,d_{2N}\}$. Then we have the proposition 3,

{\it Proposition 3}. The dispersion matrix $\sig(\vec{Q},\rho )$ can be
diagonalized by means of linear CT of coordinates and momenta,

$$
\vec{Q}\rar \vec{Q}_{d} = \Lam_{d}\vec{Q}:\quad \sig(\vec{Q},\rho )
\rar  \sig(Q_{d},\rho) = {\rm diag}\{d_{1},d_{2},\ld
,d_{2N}\}.       \eqno{(29)}$$
 To prove the diagonalization formula (29) we  have  (in  view
of  the positive definiteness of $\sig(\vec{Q},\rho )$ and the result of
ref. [22])  only  to note the following  transformation law of  the
dispersion  matrix  under linear CT $\vec{Q}^\pr  = \Lam\vec{Q}$,
$$
\sig(Q^\pr ,\rho ) = \Lam \sig(\vec{Q},\rho )\Lam ^{T},
\eqno{(30)}$$
 which is easily derived from eqs. (29) and the definition  (20)
of  $\sig$ .
Note that: (a) the diagonalizing CT $\Lam_{d}$ is not unique [25];\,\,
(b) $\Lam_d$ is state dependent: $\Lam _{d} = \Lam _{d}(\rho )$.
Proposition 3 means that any quantum state is unitary and methaplectically
equivalent to a state with vanishing covariances of coordinates and
momenta. The general methaplectic operator (14) is called also squeeze
operator [7,8,11,30], its canonical one dimensional form being
$\exp[(z^*a^2-za^{\dg 2})/2]$. It is more adequate to call it {\it squeeze and
correlation operator} since, e.g., for pure imaginary $z$ it generates
covariance of $p$ and $q$ and dosn't squeeze, while for real $z$ it
generates squeezing and doesn't correlate.   

The second property of $\sig(\vec{Q},\rho)$ which we note is that its
determinant doesn't depend on the linear CT as it is  seen from eq. (30),
where $\det \Lam  = 1$,
$$
\det\sig(Q^\pr,\rho) \equiv \det\sig(\Lam Q,\rho)=\det\sig(\vec{Q},\rho ).
\eqno{(31)}$$
The trace of $\sig$ is generally not invariant, however the quantities
Tr$[(\sig J)^{2k}],\,\,k=1,2,\ld ,$ are invariant. Recall that time
evolution governed by quadratic Hamiltonians is equivalent to linear CT of
$\vec{p},\,\vec{q}$.

The  third  property  of $\sig(\vec{Q},\rho )$  to  note  is  referred   to
the minimization of the Robertson inequality (27). We formulate  it  as  a
proposition.

{\it Proposition 4}. The equality in the multimode Robertson uncertainty
relation, eq.(27), holds in pure  states $|\Psi \ra$  only  and  iff
$|\Psi \ra$  is unitary equivalent to a multimode CCS $|\vec{\alf}\ra$,
$$
|\Psi \ra = U(\Lam )|\vec{\alf}\ra,
\eqno{(32)}$$
 where $U(\Lam)$ is an operator of the methaplectic group $Mp(N,R)$.

{\it Proof}. Let us first recall that $Mp(N,R)$ is a  quantum  analog  of
the symplectic group $Sp(N,R) \ni  \Lam $ and  is  generated  by  the
set  of operators $\{q_{k}q_{l}, {1\over 2}(q_{k}p_{l}+ p_{l}q_{k}),
p_{k}p_{l}\}$. Thus, $U(\Lam )$  in  eq. (32)  is  an
exponent of a quadratic form of coordinates and momenta,
$$
U(\Lam) = \exp [-{i\over \hbar}(\vec{Q}\c H\vec{Q})],  \eqno{(32a)}
$$
 where $\c H$ is an $2N\times 2N$ matrix of the form (12b),
related to $\Lam $ through the formula of CT $\Lam\vec{Q}= U\vec{Q}U^\dg$.

{}For further proof we need to diagonalize the  dispersion  matrix
$\sig(\vec{Q},\rho )$  by  means  of
linear CT (29) and take into account eqs. (23a). We obtain
$$
\sig(Q_{d},\rho ) = \Lam _{d}\sig(\vec{Q},\rho )\Lam ^{T}_{d} =
\sig(\vec{Q},\rho ^\pr ), \rho ^\pr  = T(\Lam _{d})\rho
T^{\dg}(\Lam _{d}),    \eqno{(33a)}
$$
$$
\sig(Q_{d},\rho ) = \sig(\vec{Q},\rho ^\pr ) = {\rm diag}\{d_{1}(\rho
^\pr ), d_{2}(\rho ^\pr ),\ld,d_{2N}(\rho ^\pr )\},   \eqno{(33b)}
$$
 where the diagonal elements $d_{\nu }(\rho ^\pr )$ are the
squared second momenta (the squared dispersions) of $q_{k}$ and $p_{l}$ in
the new state $\rho ^\pr  = T(\Lam_{d})\rho T^{\dg}(\Lam_{d})$,
$$
d_{k}(\rho^\pr) = \sig_{kk}(\rho^\pr) = {\rm Tr}[(p_{k}-\la p_{k}\ra)^{2}
\rho^\pr] \equiv  \sig_{p_kp_k}(\rho^\pr),   \eqno{(34a)}
$$
$$
d_{N+k}(\rho^\pr) = \sig_{N+k}(\rho^\pr) = {\rm Tr}[(q_{k}-\la q_{k}\ra)^{2}
\rho^\pr] \equiv  \sig_{q_k q_k}(\rho^\pr),   \eqno{(34b)}
$$
 Now we can write the determinant as  a  product  of  diagonal
element $d_{\nu}(\rho^\pr)$, $\nu = 1,2,\ld,2N$,
$$
\det\sig(\vec{Q},\rho) =\det\sig(\vec{Q},\rho^\pr) = [d_{1}(\rho^\pr)
d_{N+1}(\rho^\pr)]\ld
[d_{N}(\rho ^\pr )d_{2N}(\rho^\pr)].
  \eqno{(35)}$$
 From Heisenberg relation we have for every factor in eq. (35)
the inequality
$$
d_{k}(\rho ^\pr )d_{N+k}(\rho^\pr) \ge  {\hbar ^{2}\over 4},\quad k =
1,\ld,N.   \eqno{(36)}
$$
 From eqs. (35), (36) and (27) we derive that the equality  in
 Robertson relation (27) holds iff the equality in eq. (36) holds for all
modes (for  every
$k = 1,2,\ld N)$. And we know that this is possible iff $\rho^\pr $
is a  pure  state, namely a multimode CCS $|\vec{\alf}\ra$:
$$
\rho ^\pr  \equiv  U(\Lam _{d})\rho U^{\dg}(\Lam_{d}) =
U(\Lam_{d})|\vec{\alf}\ra \la \vec{\alf}|U^{\dg}(\Lam_{d}).
\eqno{(37)}$$
 This ends the proof of proposition 4.

The Proposition 4 can be reformulated as follows: The equality  in the
multimode Robertson  uncertainty  relation  (27)  holds  for  pure states
$|\Psi \ra$ only and iff $|\Psi \ra$ is eigenstate of a set of
boson  operators $b_{k}, k = 1,2,\ld ,N$, which are linear combination
of the original $a_{k}$ and $a^{\dg}_{k},\quad a_{k} = (2\hbar m_{k}\omega
_{k})^{-1/2}(m_{k}\omega _{k}q_{k}+ ip_{k})$, where $m_{k}$  and $\omega
_{k}$  are  mass  and frequency parameters.

Indeed, CCS $|\vec{\alf}\ra$ is an eigenstate of all $a_{k}$
with eigenvalues $\alf_{k}\,\, (a_{k}|\vec{\alf}\ra = \alf_{k}|
\vec{\alf}\ra$), and the minimizing states $|\Psi_{\rm min}\ra$ are unitary
equivalent to the CCS, $|\Psi_{\rm min}\ra =
U(\Lam_{d})|\vec{\alf}\ra$.  Then, using (37) and the BCH formula,  we
see  that $|\Psi_{\rm min} \ra$  is eigenstate of $b_{k}$,
$$    
b_{k} = U(\Lam _{d})a_{k}U^{\dg}(\Lam _{d}) = u_{kl}a_{l} +
v_{kl}a^{\dg}_{l},   \eqno{(38)}
$$
$$
b_{k}|\Psi_{\rm min} \ra = \alf_{k}|\Psi_{\rm min} \ra,\eqno{(39)}
$$
 with the same eigenvalues $\alf_{k}$. In view of (38) these
eigenstates  can be denoted as $|\vec{\alf};u,v\ra$ or $|\vec{\alf}
;\Lam \ra$. 

For quadratic Hamiltonians the time evolution operator $U_{\rm quad}(t)\in
Mp(N,R)$.  Therefore the time evolution of $|\Psi_{\rm min}\ra$ for quadratic
Hamiltonians is stable, i.e., $U(t)|\Psi_{\rm min}\ra = |\Psi^\pr_{\rm
min}(t)\ra$ up to a phase factor. $|\Psi^\pr_{\rm min}(t)\ra$ is
eigenstate of linear integrals of motions $b^0_k(t)$ with constant
eigenvalues $\alf_k$.  Overcomplete system of such eigenstates
$|\vec{\alf};\Lam (t)\ra$ (denoted as $|\vec{\alf},t\ra$) for
$N$ dimensional quadratic systems has been constructed in ref. [5] and
used  later  in many papers [10]. In ref. [26] the states $|0;\Lam \ra =
U(\Lam )|0\ra$ (where $|0\ra$ is annihilated by all boson operators
$a_{k})$ are studied  as $Sp(N,R)$ CS with maximal symmetry and in ref.
[30] the states $|\vec{\alf};\Lam\ra$ have  been further studied  and
called  multimode  squeezed  states.  States $|\vec{\alf};\Lam\ra$ which
minimize Robertson inequality should be called Robertson intelligent
states (IS). They are of the form of squeezed CS.

Let  us  note  that  in  Hilbert  space   of   pure   states   the
representation $U(\Lam )$ doesn't act transitively and all states fall in
$Mp(N,R)$ orbits, which either coincide or have no common vectors.  For
example the two orbits $U(\Lam )|\vec{\alf}\ra$ and $U(\Lam
)|\vec{\alf}^\pr \ra$ with $\vec{\alf}^\pr  \neq \vec{\alf}$ are
different.  Now it is clear that due to  the  invariance of 
det$\sig$ and Tr$[(\sig J)^{2k}]$ under the linear CT, all  states  can
be separated into larger classes with constant det$\sig$ and Tr$[(\sig
J)^{2k}]$. For example the above two orbits $U(\Lam )|\vec{\alf}\ra$ and
$U(\Lam )|\vec{\alf}^\pr \ra$ fall into one class with
$\det\sig(\vec{Q},\Psi ) = (\hbar /4)^{N}$.

A further property of the dispersion  matrix $\sig(\vec{Q},\rho)$ (the
fourth one) we want to note here is referred to its symplectic character
for a certain class of states. The matrix $\sig(\vec{Q},\rho )$ is
symplectic itself if  it is symplectic for some canonical variables
$\vec{Q}^\pr$. This follows from  the transformation formula (30). When
$\sig(\vec{Q},\rho )$ is symplectic (more precisely it is
$\sig(\vec{Q},\rho)\, (\det\sig(\vec{Q},\rho))^{-1/2N}$ which is
symplectic) it  obeys  the defining relation (19a),
$$
[\sig(\vec{Q},\rho)J\,\sig^{T}(\vec{Q},\rho)]\,[\det\sig(\rho
)]^{-1/N} = J,         \eqno{(40)}
$$
 since always $\det\sig(\vec{Q},\rho) > 1$. In terms  of  the  four
$N\times N$ blocks $\sig_{pp}(\rho)$, $\sig_{qq}(\rho)$, $\sig_{pq}(\rho)$
and $\sig_{qp}(\rho) = \sig^{T}_{pq}(\rho)$,
$$
\sig(\vec{Q},\rho) = \matb[{\sig_{pp}(\rho)} {\sig_{pq}(\rho)}
{\sig_{qp}(\rho)} {\sig_{qq}(\rho)}] \equiv
\matb[{\sig_{1}} {\sig_{2}} {\sig_{3}} {\sig_{4}}],  \eqno{(41)}
$$
the symplectic conditions (40) are rewritten as ($I$ is $N\times N$ unit
matrix)
$$
[\sig_{qq}(\rho)\sig_{pp}(\rho) - \sig^{2}_{pq}(\rho)]\,[\det
\sig(\rho)]^{-1/N} = I,     \eqno{(42a)}
$$
$$
\sig_{4}\sig_{2} - \sig^{T}_{2}\sig_{4} = 0,\quad \sig_{1}\sig_{3} -
\sig^{T}_{3}\sig_{1} = 0.   \eqno{(42b)}
$$
 Condition (42a) means that (if $\sig(\vec{Q},\rho)$ is
symplectic) the $N\times N$ matrix $\sig_{qq}(\rho)\sig_{pp}(\rho ) -
\sig^{2}_{pq}(\rho)$  is a multiple of unity, the common factor being
greater  or equal to $(\hbar ^{2}/4)^N$ (in view of relation  (27)).

We shall  prove that relations (42a,b) hold in some of the $N$ mode
squeezed  number states $U(\Lam)|\vec{n}\ra$ (with $\det\sig >
(\hbar^2/4)^N$ if $\vec{n} \neq 0$) and in all squeezed CS (Robertson IS)
$|\vec{\alf};\Lam \ra$ (with $\det\sig = (\hbar^2/4)^N$).  In the one
dimensional case the states $|\alf;\Lam \ra \equiv |\alf;u,v\ra$  are  the
Yuen two photon CS [32] and the relation (42a) is  the  equality  in  the
Robertson-Schr\"odinger  uncertainty  relation: for $N = 1$   the
quantities $\sig_{pp}(\rho)$ and $\sig_{qq}(\rho)$ are the squared
variances of $p$  and $q$  and $\sig_{pq}(\rho)$ is their covariance.

Before proceeding with the proof of (40) for the squeezed CS and Fock
states let us note that one can use symplectic conditions (42a) to define
a larger class of generalized multimode IS  (or  generalized multimode
SS), which contains canonical IS $|\vec{\alf};\Lam\ra$ as a special
case, corresponding to $\det\sig = (\hbar /4)^{N}$. This can  be  called
the  class  of  SS with symplectic uncertainty matrix.

If  one  consider  the  transformation properties of the $N\times
N$ matrix combinations occurring in (42a,b), $\sig_{4}\sig_{1} -
\sig^{T}_{2}\sig_{3},\quad  \sig_{1}\sig_{4} - \sig_{2}\sig^{T}_{3},\quad
\sig_{4}\sig_{2} - \sig^{T}_{2}\sig_{4}$ and $\sig_{1}\sig_{3} -
\sig^{T}_{3}\sig_{1}$,  one  could  see  that
they are transforming linearly  among  themselves  and  the  relations
(42a,b) are invariant under linear CT. Then it is most simple to check
the relations (42a,b) in the  diagonal  representation (33b)  of  the
dispersion matrix, where they  are  reduced  to  one  relation  (since
$\sig_{2}(\vec{Q},\rho ^\pr ) = 0 = \sig_{3}(\vec{Q},\rho ^\pr ))$.
$$
\sig_{qq}(\rho^\pr)\sig_{pp}(\rho^\pr)/[\det \sig(\rho)]^{1/N} = I.
\eqno{(43a)}
$$
 In  the  diagonal  representation $\det\, \sig(\vec{Q},\rho
^\pr) = \det [\sig_{qq}(\rho^\pr) \sig_{pp}(\rho^\pr)]$.
Eq. (43a) requires the $N\times N$ matrix $\sig_{qq}(\rho^\pr)
\sig_{pp}(\rho^\pr)$  to  be  a  multiple  of unity, i.e.
$$
\sig_{k}(\rho ^\pr )\sig_{N}(\rho ^\pr ) = f(\rho ^\pr ) \quad
{\rm for\,\,\, all }\quad k = 1,2,\ld,N.      \eqno{(43b)}
$$
 If all products $\sig_{kk}(\rho ^\pr )
\sig_{N+k}(\rho ^\pr )$ are equal then  the  common  factor
$f(\rho ^\pr )$ must be invariant under linear CT and equal to
$[\det\sig(\rho )]^{1/N}$ .

$\sig(\vec{Q})$ is diagonal in CCS $|\vec{\alf}\ra$ and in Fock states
$|\vec{n}\ra$, and eq. (43b) is satisfied in all CCS and in some Fock
states.  For  Fock  states the  matrix $\sig(\vec{Q},\vec{n})$
is  diagonal  with uncertainty products
$$
\sig_{kk}(\vec{Q},\vec{n})\sig_{N+k}(\vec{Q},\vec{n}) = {\hbar \over
4}^{2}(1 + 2n_k)^2.   \eqno{(44)}
$$
Thereby if $n_{k}$ are equal then the matrix $\sig_{qq} (\vec{Q},\vec{n})
\sig_{pp}(\vec{Q},\vec{n})$ is  a  multiple of unity with factor
$$
f(\vec{n}) = (\hbar ^{2}/4)(1 + 2n_k)^2 =
[\det\sig(\vec{Q},\vec{n})]^{1/N}.   \eqno{(44a)}
$$
 Thus, the two families of states $U(\Lam )|\vec{\alf}\ra$
and $U(\Lam )|\vec{n}\ra$ with $n_{k}= n$ obey eqs. (42a,b) and are
states with symplectic dispersion matrix. Quadratic Hamiltonians (and only
they)  will  preserve  this  symplectic property  in  time evolution. One
can easily show that eq. (43b)  can  be also satisfied in 
states $|\Psi ^\pr \ra$,  $|\Psi ^\pr \ra
= \prod_k |\vphi ^\pr\ra_{k}$, where $|\vphi ^\pr\ra_{k}$  are with
vanishing covariance: $\sig_{p_kq_k}(|\vphi^\pr\ra_k) = 0$.\\[1mm]

{\sm P.S. After the paper was completed (in August 1995) the author has
learned about article [38], where the diagonalization of the dispersion
matrix  is established by means of several (explicit) linear CT. Some
improvements in text are made in this e-print submission}.
\vs{1cm}

\leftmargin 1cm
\bc
{\bf R e f e r e n c e s}
\ec
\vs{5mm}

\hs{-1.5cm}[1]\, N.N. Bogoliubov, J. Phys. USSR    {\bf11} 23 (1947);    Zh.
Exp.  Teor. Fiz. {\bf 34} 58 (1958);  N. Cimento {\bf 7} 794 (1958); \,\,
J.G. Valatin, N. Cimento {\bf 7} 843 (1958).\\[-2mm]

\hs{-1.5cm}[2]\, K.B. Wolf, J. Math. Phys. {\bf15} 1295 (1974).\\[-2mm]

\hs{-1.6cm} [3]\, P. Kramer,\, M. Moshinsky and T.H. Seligman, in {\it Group
Theory and  its Applications}, ed. E.M. Loebl (Academic, N.Y., 1975), p.
249.\\[-2mm]

\hs{-1.5cm}[4]\, I.A. Malkin, V.I. Man'ko and D.A. Trifonov,  Phys.  Lett.
A{\bf 30}(7) 414 (1969);\,\, Phys. Rev. D{\bf 2}(8) 1371 (1970).\\[-2mm]

\hs{-1.5cm}[5]\, I.A. Malkin,   V.I. Man'ko and D.A. Trifonov,  N. Cimento A
{\bf 4}(4) 773 (1971);\,\, J. Math. Phys. {\bf 14}(5) 576 (1973).\\[-2mm]

\hs{-1.5cm}[6]\, R. Simon, E.C. Sudarshan  and N. Mukunda,  Phys. Rev. A
{\bf 37}(8) 3028 (1987). "Gaussian pure states in  quantum  mechanics and
the symplectic  group".\\[-2mm]

\hs{-1.5cm}[7]\, D.F. Walls, Nature (London) {\bf 306} 141 (1983).\\[-2mm]

\hs{-1.5cm}[8]\, R. Loudon and P.L. Knight, J. Mod. Opt. {\bf34} 709 (1987)
.\\[-2mm]

\hs{-1.5cm}[9]\, W.M. Zhang, D.H. Feng and R. Gilmore, Rev. Mod. Phys.
{\bf 62}(4) 867 (1990). "Coherent states: Theory and applications".\\[-2mm]

\hs{-1.5cm}[10]\, V.V. Dodonov  and  V.I. Man'ko,  Trudy  FIAN,  v. 183,
p.p. 1-100, 101-200 (M., "Nauka", 1987) (Proc.  of  P.N. Lebedev  Phys.
Inst., v. 183, Nuova Science, Commack, N.Y., 1989).\\[-2mm]

\hs{-1.5cm}[11]\, B.L. Schumaker, Phys. Rep. {\bf 135} 317 (1986).\\[-2mm]

\hs{-1.5cm}[12]\, R.G. Littlejohn, Phys. Rep. {\bf 138} 193 (1986).\\[-2mm]

\hs{-1.5cm}[13]\, V.V. Dodonov, O.V. Man'ko and V.I. Man'ko,  Phys. Rev. A
{\bf 50} 813 (1994). "Multidimensional Hermite polynomials and photon
distribution for  polymode mixed light";

V.V. Dodonov, V.I. Man'ko and D.E. Nikonov, Phys. Rev. A{\bf 51}(4),
3328 (1995). "Even  and  odd  coherent  states  for   multimode
parametric systems".\\[-2mm]

\hs{-1.5cm}[14]\, A.N. Seleznyova, Phys. Rev. A{\bf 51}(2) 950 (1995).
"Unitary transformations for the time-dependent quantum oscillator".\\[-2mm]

\hs{-1.5cm}[15]\, A.F.R.  de  Toledo  Piza,   Phys. Rev.  A{\bf 51}(2) 1612
(1995).  "Classical equations for quantum squeezing and  coherent  pumping
by the time-dependent quadratic Hamiltonian".\\[-2mm]

\hs{-1.5cm}[16]\, J.N. Holenhorst, Phys. Rev. D{\bf 19} 1669 (1979).\\[-2mm]

\hs{-1.5cm}[17]\, D.A. Trifonov and V.S. Gerdjikov,  Compt. Rendus  Acad.
Bulg.  Science {\bf 30}(4), 503 (1977). "Normal coordinates  for
nonstationary quantum systems".\\[-2mm]

\hs{-1.5cm}[18]\, P.G. Leach, J. Math. Phys.  {\bf 19}(2) 446 (1978).
"Quadratic Hamiltonians, quadratic invariants  and  the  symmetry group
$SU(n)$".\\[-2mm]

\hs{-1.5cm}[19]\, P.K. Colgrave  and  M.S. Abdalla,  J. Phys. A{\bf 15}
(1982) 1549.  "Harmonic oscillator with strongly pulsating mass".\\[-2mm]

\hs{-1.5cm}[20]\, W. Sarlet, J. Phys. A{\bf 11}(4) 843 (1978). "Exact
invariants for time-dependent Hamiltonian systems with one degree  of
freedom".\\[-2mm]

\hs{-1.5cm}[21]\, H. Lee and W.S. l'Yi, Phys. Rev. A{\bf 51}(2) 982 (1995).
"Non-Hermitian techniques of canonical transformations in quantum
mechanics".\\[-2mm]

\hs{-1.5cm}[22]\, J.H. Colpa,  Physica A{\bf 93} 327 (1978).
"Diagonalization of quadratic  boson  Hamiltonians";\,\,  J. Phys. A{\bf
12}(4) 469 (1979).  "Diagonalization of quadratic fermion Hamiltonians"
.\\[-2mm]

\hs{-1.5cm}[23]\, R. Bogdanovic and M.S. Gopinathan, J. Phys. A {\bf12}(9)
1457 (1979).
"A canonical  transformation  of  the   Hamiltonians   quadratic
in coordinate and momentum operators".\\[-2mm]

\hs{-1.5cm}[24]\, Y. Tikoshinsky,  J. Math. Phys.  {\bf 20}(3),  406 (1979).
"On the diagonalization of the general quadratic Hamiltonian  for  coupled
harmonic oscillators".\\[-2mm]

\hs{-1.5cm}[25]\,  L. Yeh, NASA  Conference  Publications  3219,   p. 383
 (Goddard Space Flight  Center,  Maryland,  1993).  "Infinite  mode
squeezed states and non-equilibrium statistical  mechanics  (phase
space picture approach)".\\[-2mm]

\hs{-1.5cm}[26]\, J. Klauder and B.S. Skagerstam. {\it Coherent  states}.
(World Scientific, Singapore, 1985).\\[-2mm]

\hs{-1.5cm}[27]\, Barut A.O. and Raszcka R. {\it Theory of Group
        Representations and Applications} (Polish Publishers, Warszawa,
        1977).\\[-2mm]

\hs{-1.5cm}[28]\, H.P. Robertson. Phys. Rev. {\bf 46}(9), 794-801
(1934). "An indeterminacy relation  for several observables and its
classical interpretation".\\[-2mm]

\hs{-1.5cm}[29]\, F.R. Gantmaher. {\it Teoria matritz}. (Nauka, Moskva,
1975). \\[-2mm]

\hs{-1.5cm}[30]\, X. Ma and W. Rhodes, Phys. Rev. A {\bf 41}(9) 4625 (1990).
 "Multimode squeeze operators and squeezed states".\\[-2mm]

\hs{-1.5cm}[31]\, D.A. Trifonov, J. Math. Phys. {\bf 35}(5) 2297
(1994).\\[-2mm]

\hs{-1.5cm}[32]\, H. Yuen,  Phys. Rev. A{\bf 13} 2226 (1976).\\[-2mm]

\hs{-1.5cm}[37]\, E. Kamke, {it Differentialgleichungen, L\"osungsmethoden
und L\"osungen}, v. 1 (Akademishe Verlagsgesellschaft, Leipzig, 1959).
 \\[-2mm]

\hs{-1.5cm}[38]\, E.C.G. Sudarshan, C.B. Chiu and G. Bhamathi, Phys. Rev.
A{\bf 52}(1) 43 (1995). "Generalized uncertainty relations and
characteristic invariants for the multimode states".
\end{document}